\documentclass[english]{article}
\usepackage{mathptmx}
\usepackage[T1]{fontenc}
\usepackage[latin9]{inputenc}
\usepackage{color}
\usepackage{amstext}
\usepackage{graphicx}

\makeatletter
\newcommand{\lyxaddress}[1]{
\par {\raggedright #1
\vspace{1.4em}
\noindent\par}
}

\usepackage{mathptmx}

\usepackage[a4paper]{geometry}

\@ifundefined{definecolor}
 {\usepackage{color}}{}

\usepackage{axodraw}

\usepackage{babel}

\@ifundefined{showcaptionsetup}{}{%
 \PassOptionsToPackage{caption=false}{subfig}}
\usepackage{subfig}
\makeatother

\usepackage{babel}

\begin{document}

\title{Possible Discovery Channel for New Charged Leptons at the LHC}

\author{V. E. Özcan$^{a}$, S. Sultansoy$^{b,c}$ and G. Ünel$^{d}$}

\maketitle

\lyxaddress{\textcolor{black}{$^{a}$} Department of Physics and Astronomy, University
College London, London, UK\textcolor{black}{.} \\
 \textcolor{black}{$^{b}$ Institute of Physics, Academy of Sciences,
Baku, Azerbaijan.} \\
 \textcolor{black}{$^{c}$ TOBB ETU, Physics Department, Ankara,
Turkey.} \\
 \textcolor{black}{$^{d}$ University of California at Irvine,
Physics and Astronomy Department, Irvine, USA.} }
\begin{abstract}
We propose a channel for the possible discovery of new charged leptons
at the Large Hadron Collider. The proposed final state contains three
same-sign leptons, making this new channel practically backgroundless.
The method is illustrated for two different cases: the four-family
Standard Model and the Grand Unified Theory based on the E6 gauge
group. An example study taking 250 GeV as the charged lepton mass
shows that in both models,  about 8 signal events can be expected
at 14~TeV center-of-mass energy with 1~fb$^{-1}$ of integrated
luminosity. Although the event yield might not be sufficient for detailed
measurements of the charged lepton properties, it would be sufficient
to claim discovery through a counting experiment. 
\end{abstract}

\section{Introduction}

In search of what comes after the Standard Model (SM), a number of
theories based on different ideas such as Supersymmetry, Extra-dimensions
and Grand Unification have been proposed \cite{R-susy,R-ED,R-GUT}.
The upcoming experiments at the LHC will test the validity of such
ideas by searching for the new interactions or the new particles (bosons
or fermions) predicted by these models. Among the most studied beyond
the SM theories predicting new fermions one can cite the 4-family
Standard Model (SM4), unified models based on E6 gauge group (E6GUT)
and Little Higgs models \cite{DMM,R-e6,little-higgs}.

The quark sector of the predicted fermions has received a lot of attention
and the discovery potential at the LHC, large due to copious QCD production,
was extensively discussed \cite{R-atlas-tdr,Holdom,biz_FF}. The models
like E6GUT and SM4 also predict new heavy leptons. Some early studies
considered the pair production of the heavy leptons mediated by $\gamma/Z/Z'$,
as a means of discovery \cite{lepton-pair}. The smallness of the
pair production cross-section, compared to the quark sector, and its
dependency on the $Z'$ mass required consideration of integrated
luminosities corresponding to many years of LHC operation. It was
also thought that the search for the new heavy leptons would benefit
from the clean environment of future lepton colliders \cite{clic-phys-tdr,clic_tjp}.
However, lepton colliders of sufficiently high center-of-mass energy
is unlikely to be operational before 2020. Therefore the discovery
potential of the LHC has to be maximally exploited. Pursuing that
goal, it was recently shown that the LHC could possibly discover the
neutral lepton and determine its nature \cite{LHC_v4,V4_jaas}. In
this paper, we propose a channel through which the new charged lepton
can also be searched for at the LHC.

\section{Proposed Discovery Channel}

\begin{figure}
\begin{centering}
\includegraphics[scale=0.5]{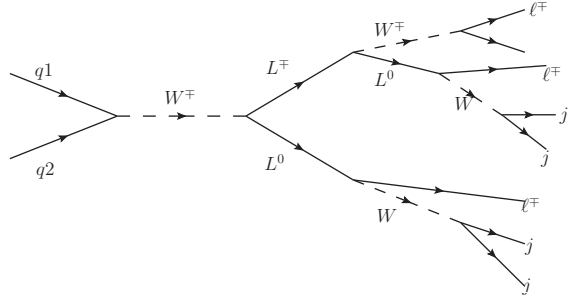} 
\par\end{centering}

\caption{The proposed production and decay of new charged leptons at the LHC}

\label{Flo:prod_decay} 
\end{figure}

We approach the problem of discovering new heavy leptons under relatively
model-independent assumptions. We will denote the generic charged
heavy lepton with $L^{\mp}$ and the neutral one with $L^{0}$. The
discovery channel we propose is $pp\rightarrow W^{\mp*}\rightarrow L^{\mp}L^{0}$,
where the charged lepton is assumed to be heavier than the neutral
one, $m_{L^{\mp}}>m_{L^{0}}$, resulting in the usual weak decay $L^{\mp}\to W^{\mp}L^{0}$,
with the $W^{\mp}$ being real or virtual. In the best discovery scenario,
the heavy neutral lepton is of Majorana type. In this case, the decays
of the neutral leptons would yield two known leptons of same sign
fifty-percent of the time: $L^{\mp}L^{0}\to W^{\mp}L^{0}L^{0}\to W^{\mp}W\ell^{\mp}W\ell^{\mp}$.
When the primary $W^{\mp}$ boson, originating from $L^{\mp}$, decays
leptonically ($e$ or $\mu$), the resulting third lepton is also
of the same sign as the first two, yielding in a final state with
three same-sign leptons. This tree level production and decay process
is illustrated in Fig.~\ref{Flo:prod_decay}. 

\begin{figure}
\begin{centering}
\includegraphics[width=0.6\paperwidth]{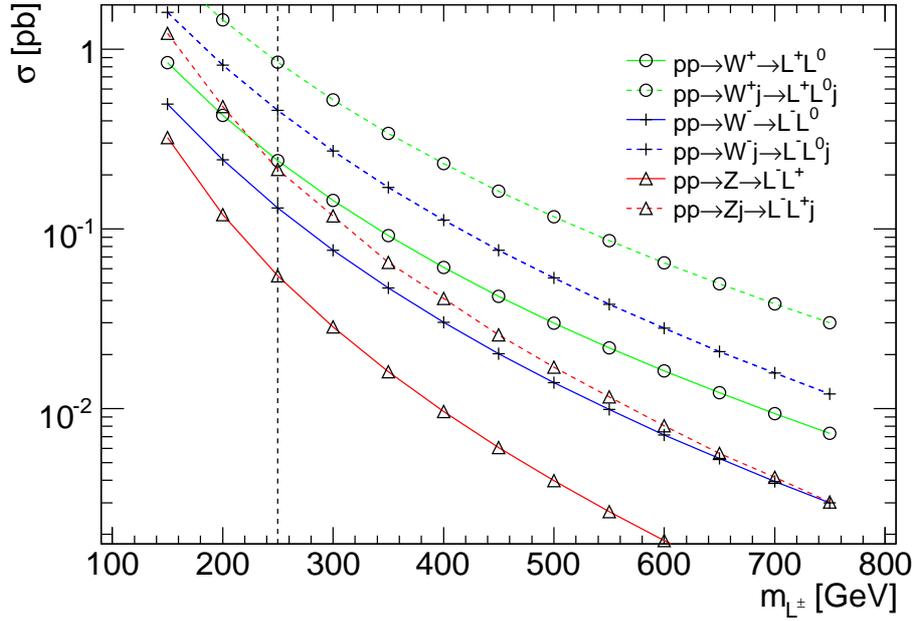} 
\par\end{centering}

\caption{Tree-level cross-sections at the LHC running at $\sqrt{s}=14$~TeV
for various production mechanisms of heavy leptons as functions of
the charged lepton mass. The mass of the neutral heavy lepton, is
taken to be $100$~GeV. The vertical dashed line represents the reference
scenario studied. }

\label{fig:xsec} 
\end{figure}

The cross-sections for the two possible $L^{\mp}$ production channels
at the tree level have been calculated using the CompHEP 4.4.3 generator~\cite{R-comphep}.
These channels are the pair production via $Z$ boson and the $L^{\mp}\; L^{0}$
associated production via the $W$ boson. The expected LHC cross-sections
are shown in Fig.~\ref{fig:xsec} as a function of the charged lepton
mass. The mass of the heavy neutrino is taken as $100$~GeV in accordance
with the recently published limits by the Particle Data Group~\cite{PDG}.
It is seen that the $Z$-mediated channel has a lower cross-section
and the consideration of the  production process with an additional
jet, shown by the dashed lines, increases the cross-sections and thus
improves the discovery prospects.

\section{Application to 4th Generation Leptons}

The imminent operation of the LHC renewed the interest in a fourth
Standard Model (SM) family \cite{4SM-SW}. The SM4 is predicted by
the so called {}``flavor democracy'' approach, which aims to explain
the currently observed fermion mass hierarchy \cite{FLDem review}.
Furthermore, it gives an opportunity to explain the baryon asymmetry
of the universe within the SM using the available CP violating phase
in the extended CKM matrix~\cite{BAU}. The 3-family SM (SM3) leads
to a ten orders of magnitude smaller CP violating phase compared to
the observations \cite{BAU}. The SM4 is obtained by extending the
SM3 with an additional set of quarks and leptons denoted as: $u_{4}$
and $d_{4}$ for quarks, $e_{4}$ for the charged lepton and $\nu_{4}$
for the heavy neutrino. The fermion-boson interaction vertices of
the fourth family fermions are similar to the first three families.
Although the masses and the mixings of the new fermions are not fixed
by the SM, they are bound by the experimental and precision fit results
\cite{Tait,okunlar,he,PDG,sher-hung}.

\begin{figure}
\begin{centering}
\includegraphics[width=0.6\paperwidth]{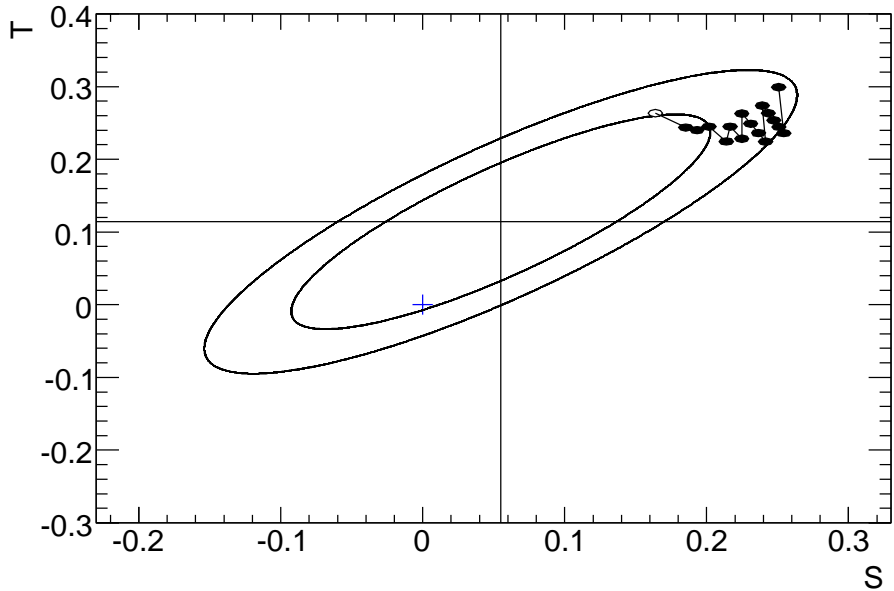}
\par\end{centering}

\caption{The 1 and 2 sigma error ellipses on the (S,T) parameters centered
on the experimental value obtained by the LEP Electroweak Working
Group. The SM, shown by the small cross, corresponds to (S=0,T=0)
calculated with $m_{top}=$170.9~GeV and $m_{H}=$115~GeV. The open
circle is the (S,T) value calculated using Majorana neutrinos of 100
and 900 GeV ($\cos^{2}\theta=0.9$), $m_{e_{4}}=250$~GeV, $m_{u_{4}}=260$~GeV,
$m_{d_{4}}=260$~GeV and $m_{H}=$115~GeV. The black circles represent
the cases for increasing Higgs mass, from 150 to 900~GeV in steps
of 50~GeV following the connecting line where the $u_{4}$ mass also
slowly increases to 330~GeV.}
\label{Flo:stu_mj}
\end{figure}

The contributions to electroweak oblique parameters S and T arising
from new fermions and the Higgs boson are calculated using the exact
one-loop calculations given in References~\cite{he,STU_peskin,STU_majo}.
As a reference scenario, we consider  Majorana-type heavy neutrinos
of mass 100 and 900~GeV, yielding a mixing angle of $\cos^{2}\theta=0.9$,
and an equivalent Dirac mass of $\nu_{D}=300$~GeV in the nomenclature
of Ref.$~$\cite{STU_majo}. The charged lepton mass is taken to be
250~GeV. For different values of the Higgs mass, a scan of the $u_{4}$
and $d_{4}$ mass values is performed to find the set yielding the
S,T values closest to the experimental value obtained by the LEP Electroweak
Working Group. The results are shown in Fig.~\ref{Flo:stu_mj}, where
the ellipses represent the 68\% and 95\% confidence level constraints
on the S and T parameters. We observe that for all Higgs masses between
115 and 900~GeV, the SM4 with Majorana neutrinos can provide solutions
within the 2 sigma error ellipse. It is interesting to note that,
for each Higgs mass value, the minimization process yields masses
for the fourth family fermions that are within about 70~GeV of each
other as predicted by the DMM model \cite{DMM}. For example for a
Higgs of 300~GeV, the masses of the SM4 quarks are found to be 290
and 260~GeV for $u_{4}$ and $d_{4}$, respectively.

\subsection{Expectations for $e_{4}$, $\nu_{4}$}

In this section, we use a parameterization of the $4\times4$ PMNS
matrix in agreement with the current neutrino oscillation data. In
Ref.~\cite{DDM-Param}, the four-dimensional CKM matrix has been
parameterized as a modification of $4\times4$ unit matrix, and the
values for the three degrees of freedom in this parameterization have
been extracted from the available experimental data on the fermion
masses. The parameterization is common between the quark and lepton
sectors and predicts $Br(e_{4}\to W\nu_{4})\simeq1$, $Br(\nu_{4}\to W\mu)=0.68$
for different values of the assumed unified Yukawa coupling coefficient
and the corresponding values of the aforementioned parameters.

$pp\rightarrow W\nu_{4}\nu_{4}\to WW\mu W\mu$ has been studied as
the particular final state where the considered decays are leptonic
for the first and hadronic for the remaining two $W$ bosons. As mentioned
previously, the two muons in the final state will be of the same sign
fifty-percent of the time due to Majorana nature of the $\nu_{4}$.
The branching fraction of these three same-sign lepton final states
is $0.68^{2}\times0.22\times0.68^{2}\times0.5^{2}\simeq1.2\%$. The
effective cross-section for this particular state is $4.44\,$fb for
$m_{e_{4}}=250\,$ GeV when no additional jets are taken into account.
The same final state with an additional jet would enhance the cross-section
by roughly a factor of four, yielding about 18  {}``backgroundless''
events per fb$^{-1}$.

\section{Application to E6 Leptons}

In the GUT (Grand Unified Theory) models, the SM gauge group $\text{SU}(3)_{C}\times\text{SU}(2)_{L}\times\text{U}(1)_{Y}$
is embedded into a larger symmetry group, which is recovered at a
higher scale. Such models predict new gauge bosons (e.g. $W'$, $Z'$)
which are sought at the current accelerators and new fermions which
may also be at the reach of the LHC. Among the most frequently studied
ones, the super-string inspired, E6GUT model can be cited. It has
a 27-plet per family which contains, among other fields, two fermion
fields, denoted as $N$, and $E^{\mp}$ that could have the quantum
numbers of a heavy neutral and charged lepton, respectively. We assume
that only the members of a single 27-plet would be light enough to
be accessible at the LHC. This assumption is pessimistic since further
accessible families would increase the production cross-section. Secondly,
we also assume the in-family mixings to be stronger than the inter-family
mixings, motivated by current experimental results, giving a quasi
diagonal CKM matrix.

The interesting properties of this model are the existence of flavour
changing neutral currents (FCNCs) and the right-handed weak currents
with vertices proportional to the mixing angle $\sin\theta_{R}$ which
are severely constrained from the current precision data. Another
severely constrained quantity is the difference between the mixing
angles of the left components of the heavy leptons to their ordinary
counterparts, $\delta\equiv\theta_{\nu_{e}}^{L}-\theta_{e}^{L}$,
which has the potential of modifying the V-A structure of the model.
We will take both of these mixings to be vanishingly small for the
rest of this note, suppressing $E^{\mp}\to W^{\mp}v$ and $E^{\mp}\to Z\, e^{\mp}$
decays. In this case, the remaining $E^{\mp}$ decay mode is via the
charged current process $E^{\mp}\to W^{\mp}N$, provided $m_{E^{\mp}}>m_{N}$
.

\subsection{Expectations for $E^{\mp}$, $N$}

After the simplifying assumptions discussed in the previous section,
the $ENW$ vertex becomes identical to the SM4 case allowing us to
use the same computations for the $E$ production at the LHC. The
Dirac or Majorana nature of the neutrinos defines the ratio of branching
fractions in the neutral current decays to its charged current decays.
For Dirac-type ordinary neutrinos, and all mixing angles taken equal,
this ratio is given as $Br(N\to Z\nu_{e})/Br(N\to We)=1/4$ \cite{R-hewet-rizzo}.
The discovery channel becomes $pp\to W^{\mp}\to E^{\mp}N\to W^{\mp}NN\to\nu_{\ell}\ell^{\mp}We^{\mp}We^{\mp}$
where $\ell$ could be an electron or a muon and the remaining two
$W$ bosons could be reconstructed using their hadronic decays. The
total branching fraction is calculated as $0.22\times0.75^{2}\times0.5^{2}\times0.68^{2}=$
1.4\% making the effective cross-section for this particular state
$5.2\,$fb for $m_{E}=250\,$ GeV without considering the additional
jets. The consideration of a single additional jet increases the expected
signal yield to about 21 events per fb.

\section{Detector Effects and Estimates of the Backgrounds}

In this section, we consider the detector acceptance for the signal
leptons and provide an order-of-magnitude estimate for backgrounds
that might arise due to detector effects. For these purposes, we refer
to the expected performance of the ATLAS Detector$~$\cite{atlas-csc}
as an example. In order to exploit both the single and di-lepton triggers
for this detector, we require all leptons to have $P_{T}>10\,$ GeV
and $|\eta|<2.5$, and at least one lepton to satisfy $P_{T}>20\,$
GeV. Since the neutrinos of the reference scenario are relatively
light, these requirements reject about half of the signal events.
The requirements have similar or lower efficiencies for the three
example background processes we discuss below, and, they are simply
folded into our calculations.

\begin{figure}
\begin{centering}
\subfloat[]{\begin{centering}
\includegraphics[scale=0.38]{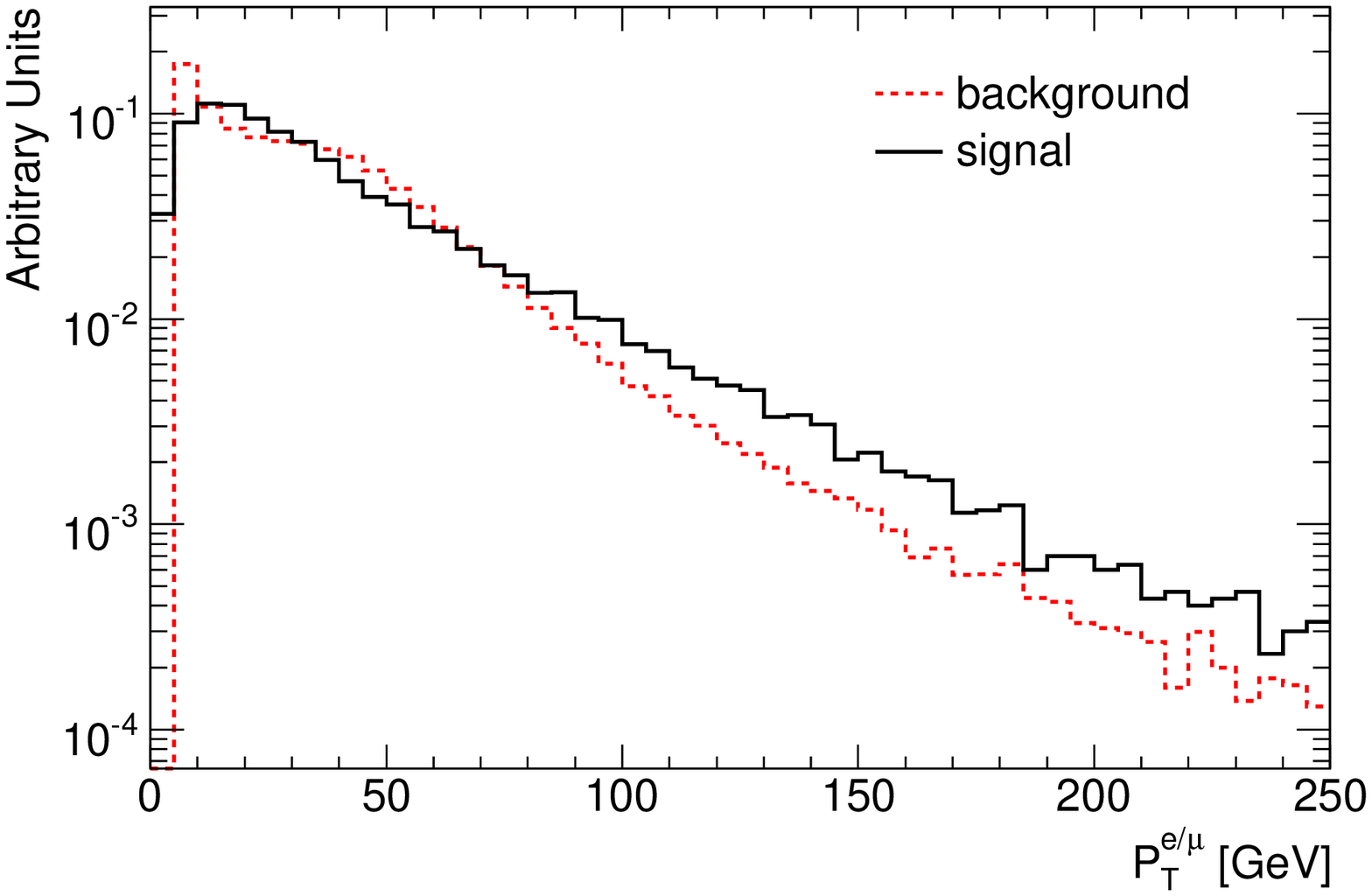}
\par\end{centering}

\label{fig:bg_leptons}}\subfloat[]{\begin{centering}
\includegraphics[scale=0.38]{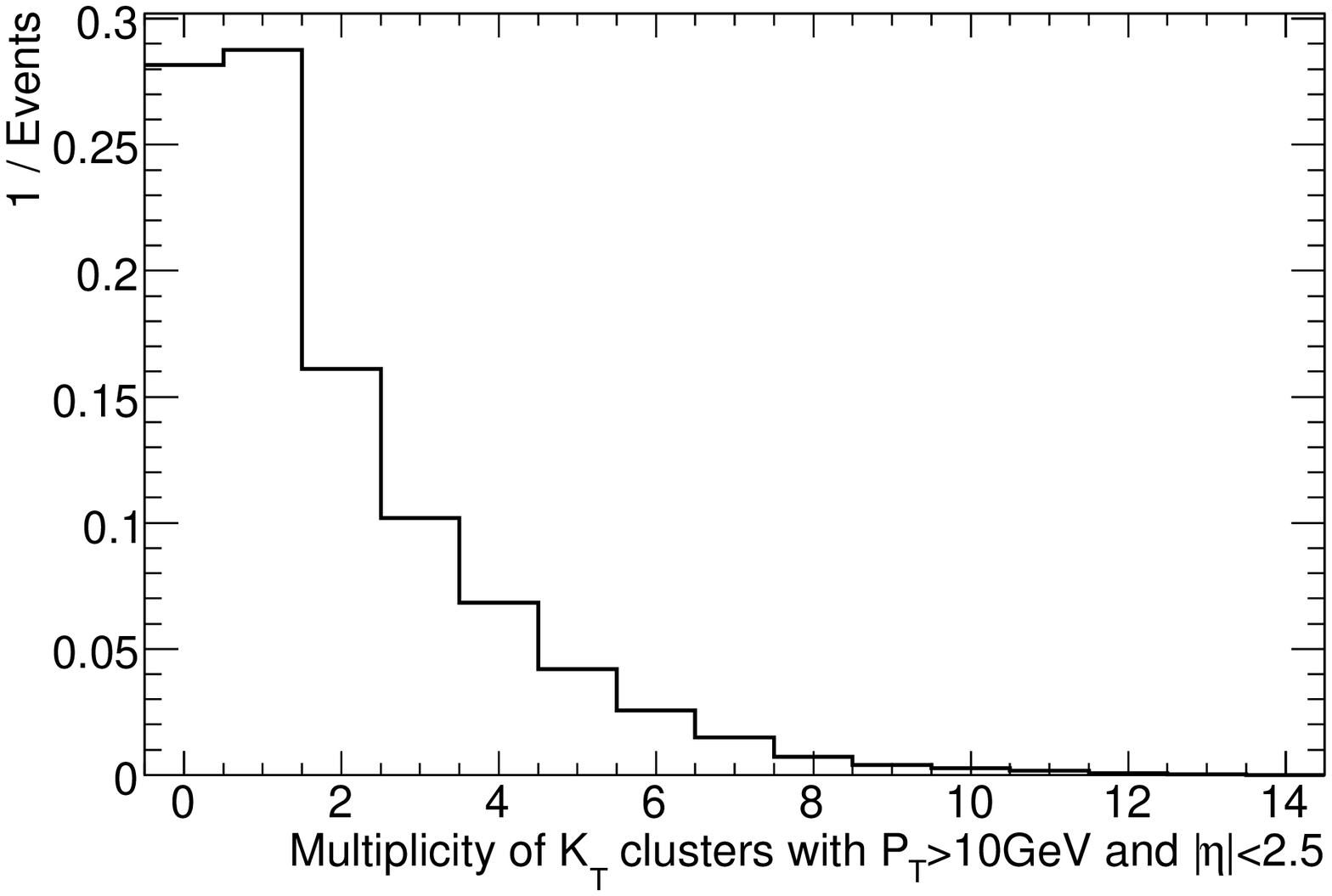}
\par\end{centering}

\centering{}\label{fig:bg_jets}}
\par\end{centering}

\begin{centering}
\subfloat[]{\centering{}\includegraphics[scale=0.38]{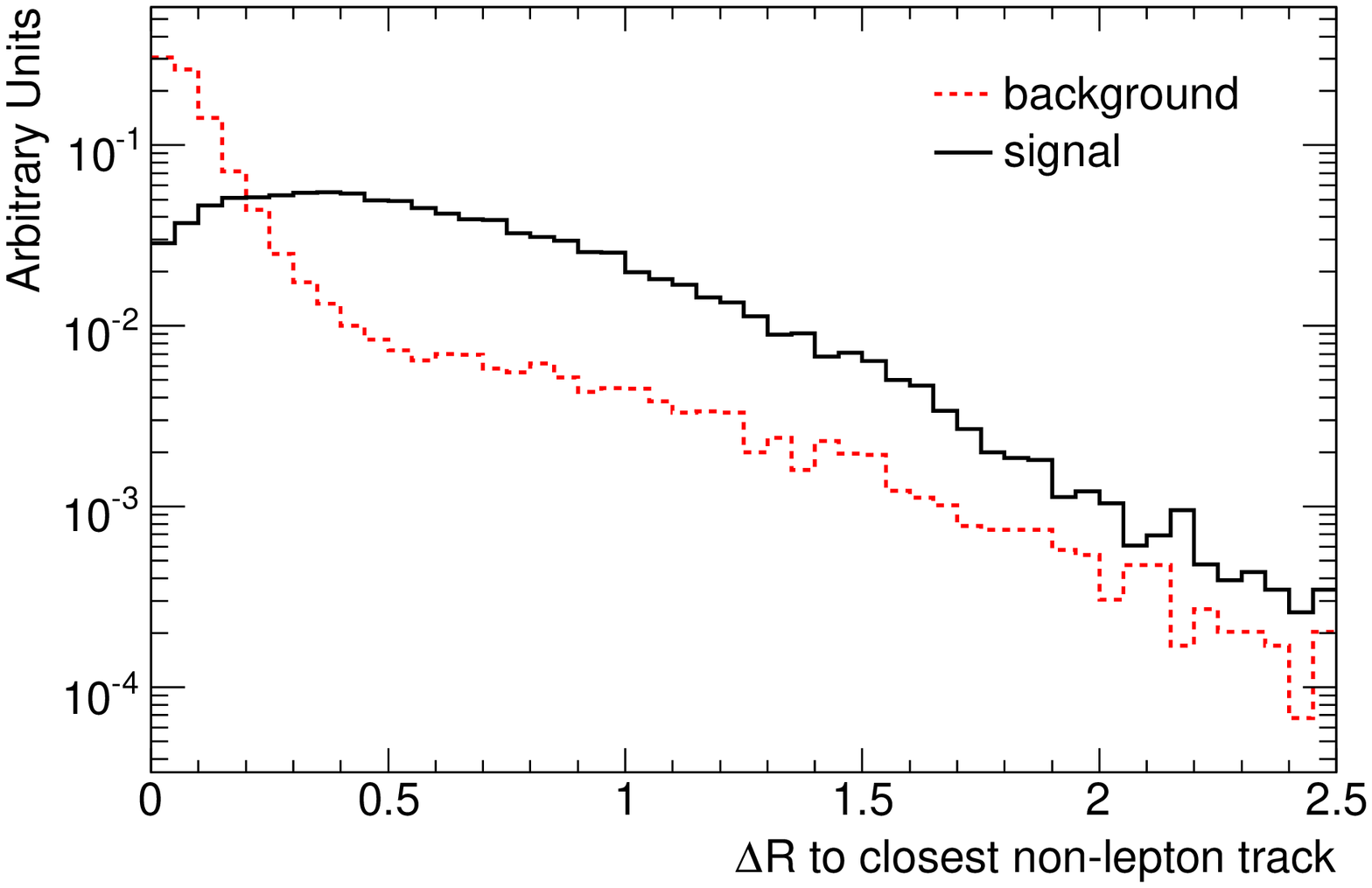}\label{fig:deltaR}}\subfloat[]{\begin{centering}
\includegraphics[scale=0.38]{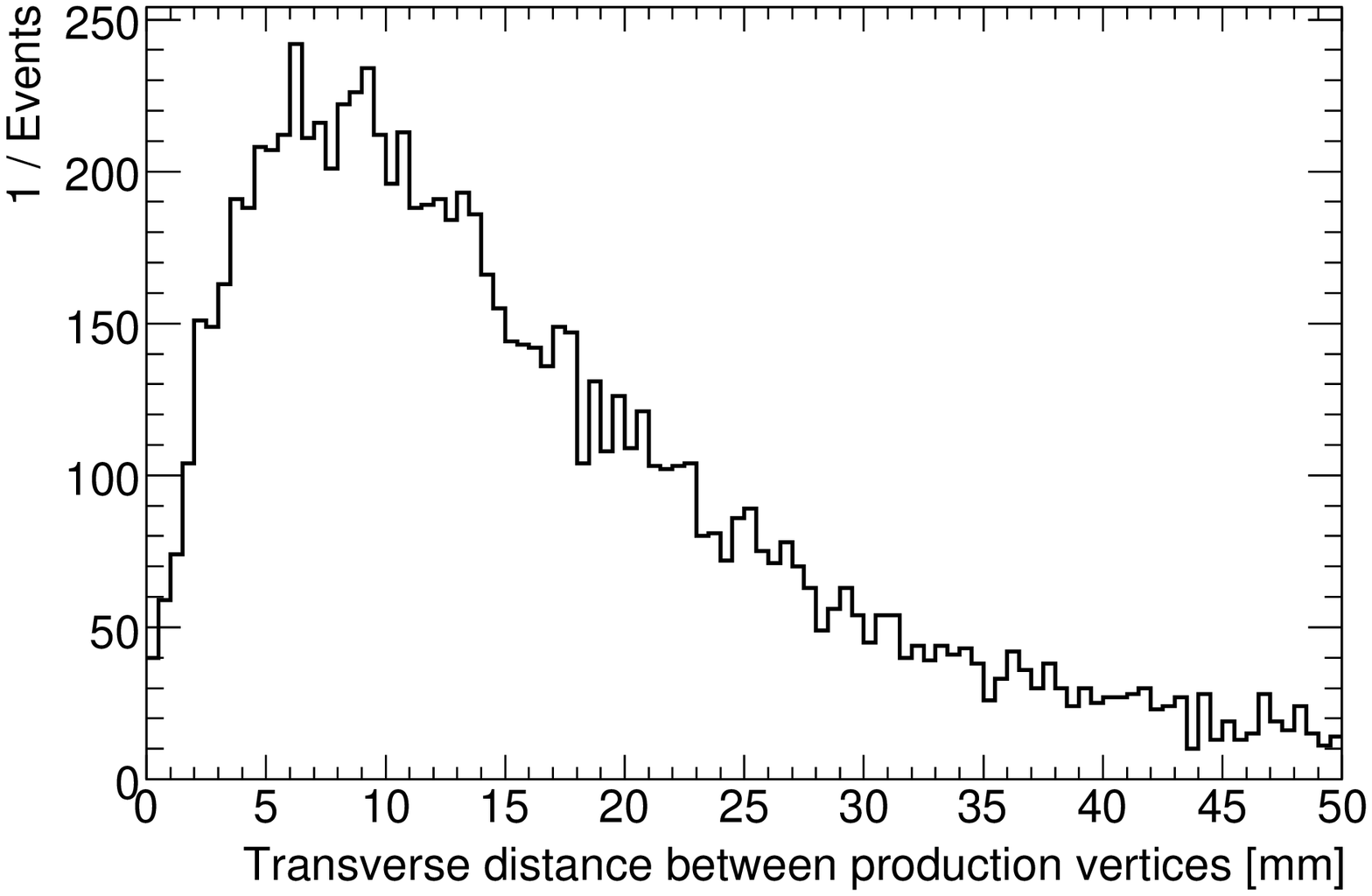}\label{fig:mtd}
\par\end{centering}

}
\par\end{centering}

\centering{}\caption{Distributions for variables relevant to the estimation of the backgrounds.
(a) Transverse momenta for charged leptons from signal and $\ell^{+}\ell^{-}\ell^{+}\nu$
and $\ell^{+}\ell^{-}\ell^{-}\bar{\nu}$ background events. (b) Multiplicity
for jets with $P_{T}>10\,$GeV in $W+jet$ events. (c) $\Delta R$
to the closest non-lepton track for leptons in the signal events and
for leptons originating from jets in $t\bar{t}$ events. (d) Maximum
transverse distance between any of the two lepton production vertices
in $t\bar{t}$ events.}

\end{figure}

First possible source of background is from SM3 processes to final
states with three charged leptons of differing signs. We generate
75 thousand such events in the $\ell^{+}\ell^{-}\ell^{+}\nu$ and
$\ell^{+}\ell^{-}\ell^{-}\bar{\nu}$ final states (where $\ell=e,\mu$
and any combination of flavors is allowed) with Madgraph/MadEvent
$~4.2.0~$\cite{madgraph}. Based on muon data in Ref.$~$\cite{atlas-csc},
charge mis-measurement rate is conservatively parameterized as $\epsilon_{mischarge}=10^{-4+P_{T}/200{\rm {GeV}}}$,
where $P_{T}$ is the lepton transverse momentum. Using this parametrization,
a pseudo Monte Carlo study on the generated Madgraph events, for which
the lepton $P_{T}$ are shown in Fig.$~$\ref{fig:bg_leptons}, yields
the probability of obtaining three same-sign leptons to be $(2\pm0.5)\times10^{-4}$.
Multiplying by the total cross-section of $195.7\pm0.6\,$fb, we find
that the expected number of background events is only $0.04$ per
fb$^{-1}$.

The second possible source of background is from misidentification
of jets as leptons. For an estimation of the size of this type of
background, we refer to the expected performance of the maximum-likelihood
based electron identification method, and take the jet rejection factor
(inverse of the probability of misidentfiying jets as leptons) as
$3.77\times10^{4}$. (The corresponding electron identification efficiency
for this level of rejection is 77\%.) To convert this rejection factor
into an estimate of the backgrounds, we generate $W+jet$ events with
Pythia~6.4.14~\cite{pythia} with multiple interactions on and the
$W$ boson decaying to $e\bar{\nu}$ or $\mu\bar{\nu}$ with $P_{T}^{lepton}>10\,$
GeV and $|\eta^{lepton}|<2.5$ . This particular type of events is
interesting due to the high cross-section, $19.9\pm0.2\,$nb, and
the presence of one true isolated lepton. On 10 thousand generated
events, we run the $k_{T}$-jet algorithm \cite{ktjet} on hadron-level
final-state particles and obtain the jet-multiplicity distribution
shown in Fig.$~$\ref{fig:bg_jets}. Using this distribution and the
mentioned jet-rejection factor, we compute the probability of having
two jets misidentified as leptons with the same sign as the true lepton
in the event. Multiplying with the cross-section, the expected effective
cross-section for this type of background is found to be $0.01\,$fb.

As the third example background source, we consider the $t\bar{t}$
production since the leptons from $b$ quark decays can the potentially
emulate the signal. We generate 5 million $t\bar{t}$ events with
Pythia 6.4.14. Lepton charge and jet misidentification probabilities,
as described above, are also incorporated. For all lepton candidates,
we apply an isolation requirement of $\Delta R>0.05$, where $\Delta R$
is the distance in the $\eta-\phi$ plane, between the lepton candidate
and the closest non-lepton track of $P_{T}>1$GeV (Fig.$~$\ref{fig:deltaR}).
For the track isolation calculations the tracking efficiency is also
folded in; defined conservatively as $\epsilon_{tracking}=0.88-\frac{0.13\times|\eta|}{2.5}$.
The three-same-sign-lepton events in $t\bar{t}$ mostly originate
from one $W$ boson decaying leptonically accompanied by at least
2 leptons from $b$-jets. To eliminate such cases, we also reject
any event which has 2 or more leptons with $\Delta R<0.2$. In order
to further reduce the background, the lepton vertex information can
also be used: the signal leptons are produced at the interaction point,
a feature not found in $t\bar{t}$ events. Therefore the 3 lepton
candidates from $t\bar{t}$ events would fail to be fit into a single
production vertex. Lacking the full detector simulation and proper
vertexing algorithms, we resort to a simple selection based on the
maximum transverse distance (MTD) between any two of the lepton production
vertices. The expected secondary vertex transverse position resolution
of the ATLAS Detector, as measured in $J/\Psi\to\mu^{+}\mu^{-}$ events,
is 0.17mm or better. The MTD in $t\bar{t}$ events are significantly
higher than this value, as shown in Fig.$~$\ref{fig:mtd}; therefore
requiring MTD$\geq$0.4mm (a distance corresponding to a value better
than 2$\sigma$) would eliminate a large portion of the $t\bar{t}$
background without any significant loss in signal events. The acceptance,
trigger and selection requirements have an overall efficiency of 40\%
on signal events, whereas only 1 out of 5 million $t\bar{t}$ events
survives. Hence, using the predicted NLO cross section of 833 pb,
we expect 0.17 events per fb$^{-1}$ from $t\bar{t}$ background.

It can be argued that the selected processes are only partially representative
of the full set of all sources of backgrounds. However, it is important
to remember that for a full analysis on real data, it will be possible
to reduce these and other backgrounds by optimizing the lepton identification
criteria, by requiring one or more leptons to have relatively high
momenta, by vetoing lepton pairs with invariant close to the nominal
$Z$ mass, by using $b$-tagging, etc. These crude estimates highlight
that the backgrounds are indeed expected to be an order of magnitude
smaller than the predicted signals.

\section{Conclusion}

It is seen that the charged leptons of the fourth SM family and the
E6GUT models can be searched using the $3\ell+\geq4j+MET$ final state
where the leptons ($\ell)$ are of the same sign. The charged leptons
are expected to be 3 same sign muons 50\% of the time and 2 same sign
muons + 1 same sign electron for the other 50\% for the first model.
For the second model electrons and muons in the final state are to
be interchanged.

Although the detector effects and the reconstruction efficiency have
not been exactly calculated in this paper, the essentially backgroundless
nature of this search channel, confirmed by the order magnitude estimates,
makes it promising for the LHC, before the operation of the corresponding
lepton colliders. For an inclusive study of our reference scenario,
taking into account the selection criteria described in the previous
section,  the first (second) model yields 7 (8) events at 14~TeV
center of mass energy with 1 fb$^{-1}$ of integrated luminosity.
Although the event yield might not be sufficient for detailed measurements
of the charged lepton properties, it is sufficient to claim discovery
through a counting experiment. For example, even if the SM backgrounds
were an order of magnitude larger than our rudimentary estimate of
0.22 events/fb, the expected significance would still be over 3.5$\sigma$
using $\sqrt{2[(s+b)\ln(1+\frac{s}{b})-s]}$ as the estimator. Moreover,
the consideration of additional charged bosons ($W')$ as the mediator
in the $s$ channel and of other decay modes of the secondary $W$
bosons can further increase the effective cross-section, reducing
the integrated luminosity needed to discover the charged heavy leptons.

\subsection*{Acknowledgments}

S.S. acknowledges the support from the Turkish State Planning Committee
under the contract DPT2006K-120470. G.Ü.'s work is supported in part
by U.S. Department of Energy Grant DE FG0291ER40679. V.E.Ö. acknowledges
financial support from the UK Science and Technology Facilities Council.


\begin{thebibliography}{31}
\bibitem{R-susy} H. P. Nilles, Phys. Rept. 110, 1 (1984) and references
therein.

\bibitem{R-ED} L. Randall, R. Sundrum, Phys. Rev. Lett. 83, 3370
(1999) and references therein.

\bibitem{R-GUT}R. Blumenhagen, S. Moster and T. Weigand, Nucl. Phys.
B \textbf{751}, 186 (2006) {[}arXiv:hep-th/0603015{]}.

\bibitem{DMM}H. Fritzsch, Phys. Lett. B \textbf{289}, 92 (1992);
A. Datta, Pramana \textbf{40}, L503 (1993); A. Celikel, A. Ciftci
and S. Sultansoy, Phys. Lett. B \textbf{342}, 257 (1995).

\bibitem{R-e6}F. Gursey, P. Ramond and P. Sikivie, Phys. Lett. B
\textbf{60,} \emph{177} (1976); F. Gursey and M. Serdaroglu, Lett.
Nuovo Cimento \textbf{21}, \emph{28} (1978).

\bibitem{little-higgs} M. Schmaltz, \textit{\emph{Nucl. Phys. Proc.
Suppl}}\textit{.} B \textbf{117}, 40 (2003).

\bibitem{R-atlas-tdr}ATLAS Detector and Physics Performance Technical
Design Report. CERN/LHCC/99-14/15 (1999), section 18.2; ; E. Arik
et.al., Phys. Rev. D \textbf{58}, 117701 (1998).

\bibitem{Holdom}B. Holdom, JHEP \textbf{0608}, 076 (2006); B. Holdom,
JHEP \textbf{0703}, 063 (2007); B. Holdom, arXiv:0705.1736 (2007).

\bibitem{biz_FF}V. E. Ozcan, S. Sultansoy and G. Unel, Eur. Phys.
J. C \textbf{57}, 621 (2008).

\bibitem{lepton-pair} C.Alexa and S.Dita, ATL-PHYS-2003-014 (2003).

\bibitem{clic-phys-tdr} E. Accomando \emph{et al.} {[}CLIC Physics
Working Group{]}, {[}arXiv:hep-ph/0412251{]}.

\bibitem{clic_tjp}R. Ciftci, A. K. Ciftci, E. Recepoglu and S. Sultansoy,
Turk. J. Phys. \textbf{27}, 179 (2003) {[}arXiv:hep-ph/0203083{]}.

\bibitem{LHC_v4}T. Cuhadar-Donszelmann, M. K. Unel, V. E. Ozcan,
S. Sultansoy and G. Unel, JHEP \textbf{0810}, 074 (2008) {[}arXiv:0806.4003
{[}hep-ph{]}{]}.

\bibitem{V4_jaas}F. del Aguila and J. A. Aguilar-Saavedra, {[}arXiv:0808.2468
{[}hep-ph{]}{]}.

\bibitem{4SM-SW}B. Holdom, W. S. Hou, T. Hurth, M. L. Mangano, S.
Sultansoy and G. Unel, {[}arXiv:0904.4698 {[}hep-ph{]}{]}.

\bibitem{FLDem review}S. Sultansoy, AIP Conf. Proc. \textbf{899},
49 (2007) {[}arXiv:hep-ph/0610279{]} and references therein.

\bibitem{BAU}G. W. S. Hou, ICHEP 08 Conf. Proc., {[}arXiv:0810.3396/hep-ph{]}.

\bibitem{Tait}G.D. Kribs, T. Plehn, M. Spannowsky and T.M.P. Tait,
Phys. Rev. D \textbf{76}, 075016, (2007) {[}arXiv:0706.3718{]}.

\bibitem{okunlar}V. A. Novikov, L. B. Okun, A. N. Rozanov and M.
I. Vysotsky, Phys. Lett. B \textbf{529}, 111 (2002) {[}arXiv:hep-ph/0111028{]};
V. A. Novikov, L. B. Okun, A. N. Rozanov and M. I. Vysotsky, JETP
Lett. \textbf{76}, 127 (2002) {[}arXiv:hep-ph/0203132{]}.

\bibitem{he}H. J. He, N. Polonsky and S. F. Su, Phys. Rev. D \textbf{64},
053004 (2001) {[}arXiv:hep-ph/0102144{]}.

\bibitem{STU_peskin}M. E. Peskin and T. Takeuchi, Phys. Rev. D \textbf{46},
381 (1992). 

\bibitem{STU_majo}B. A. Kniehl and H. G. Kohrs, Phys. Rev. D \textbf{48},
225 (1993).

\bibitem{PDG}C. Amsler et al., Phys. Lett. B\textbf{~667}, 1 (2008).

\bibitem{sher-hung}P.Q. Hung and Mark Sher, Phys. Rev. D \textbf{77},
037302 (2008) arXiv:0711.4353 {[}hep-ph{]}.

\bibitem{DDM-Param}A. K. Ciftci, R. Ciftci and S. Sultansoy, Phys.
Rev. D \textbf{72} 053006 (2005).

\bibitem{R-comphep}A. Pukhov, arXiv:hep-ph/0412191 (2004); E. Boos
et al. (CompHEP Collaboration), Nucl. Instrum. Meth. A \textbf{534},
\emph{250} (2004).

\bibitem{R-hewet-rizzo}J. Hewett and T. Rizzo, Phys. Rep. \textbf{183},
193 (1989). 

\bibitem{atlas-csc}ATLAS Collaboration, \emph{Expected Performance
of the ATLAS Experiment, Detector, Trigger and Physics}, CERN-OPEN-2008-020
(2009).

\bibitem{madgraph}J. Alwall et al., \emph{MadGraph/MadEvent v4: The
New Web Generation, }JHEP \textbf{0709} (2007) 028.

\bibitem{pythia}T. Sj{ö}strand, S. Mrenna and P. Skands, \emph{PYTHIA
6.4 Physics and Manual}, JHEP \textbf{0605} (2006) 026.

\bibitem{ktjet}S. Catani, Yu. L. Dokshitzer, M. H. Seymour and B.
R. Webber, Nucl. Phys. B\textbf{ 406} (1993) 187.
\end{thebibliography}
\end{document}